\documentclass[12pt]{article}
\usepackage{cite,epsfig,amssymb,amsmath,graphicx,color}
\allowdisplaybreaks

\evensidemargin \oddsidemargin

\newcommand{\nn}{\nonumber}

 \def\rh{\rho}

\def\nn{\nonumber}

\begin{document}

\begin{titlepage}

%-------------------- footnote symbol in title page -----------------
\renewcommand{\thefootnote}{\fnsymbol{footnote}}

%----------------------- preprint number & date ---------------------

\begin{flushright}
KIAS-P14042
\end{flushright}

%---------------------------- Title ---------------------------------
\vspace{15mm}
\baselineskip 9mm
\begin{center}
  {\Large \bf Holographic Entanglement Entropy of \\
  Mass-deformed ABJM Theory}
\end{center}

%--------------------- Authors and Addresses ------------------------
\baselineskip 6mm
\vspace{10mm}
\begin{center}
 Kyung Kiu Kim,$^1$ O-Kab Kwon,$^2$ Chanyong Park,$^2$
 and Hyeonjoon Shin$^3$
 \\[10mm]
  $^1${\sl Department of Physics and Photon Science,
   School of Physics and Chemistry, \\ GIST, Gwangju 500-712, Korea}
  \\[3mm]
  $^2${\sl Institute for the Early Universe, Ewha Womans University,
  Seoul 120-750, Korea}
  \\[3mm]
  $^3${\sl School of Physics, Korea Institute for Advanced Study,
  Seoul 130-722, Korea}
  \\[10mm]
  {\tt kimkyungkiu@gmail.com,~okabkwon@ewha.ac.kr,
cyong21@ewha.ac.kr,~hyeonjoon@kias.re.kr}
\end{center}

\thispagestyle{empty}

%-------------------------- abstract --------------------------------
\vfill
\begin{center}
{\bf Abstract}
\end{center}
\noindent
We investigate the effect of supersymmetry preserving mass deformation
near the UV fixed point represented by the ${\cal N}=6$ ABJM theory.
In the context of the gauge/gravity duality, we analytically calculate
the leading small mass effect on the renormalized entanglement entropy (REE)
for the most general Lin-Lunin-Maldacena (LLM) geometries
in the cases of the strip and disk shaped entangling surfaces.
Our result shows that the properties of the REE in (2+1)-dimensions are consistent
with those of the $c$-function in (1+1)-dimensions.
We also discuss the validity of our computations in terms of the curvature
behavior of the LLM geometry in the large $N$ limit and the relation between
the correlation length and the mass parameter for a special LLM solution.
\\ [3mm]


\end{titlepage}

\baselineskip 6.6mm
\renewcommand{\thefootnote}{\arabic{footnote}}
\setcounter{footnote}{0}

\tableofcontents

\section{Introduction}

The entanglement entropy (EE) has become an important quantity in a
wide range of research areas, from condensed matter physics to quantum
gravity.  In quantum field theory, one of its well-known features is
the appearance of the area law describing short range correlation in
the vicinity of the boundary of two subsystems.  This correlation
causes the ultra-violet (UV) divergence in the continuum limit, which can
be regulated in terms of the UV
cutoff~\cite{Bombelli:1986rw,Srednicki:1993im}.  This implies that the
EE is a UV sensitive quantity.  However, the EE also includes some UV
insensitive information for the degrees of freedom related to the long
range correlations of the system.  One important task for further
exploration of such long range degrees of freedom is to define an
appropriate finite quantity in the continuum limit, analogous to the
Zamolodchikov $c$-function in 2-dimensional quantum field
theory~\cite{Zamolodchikov:1986gt}.

%Except for the physics related to the UV cutoff, however, the EE also
%includes some information for long range correlations related to the
%degrees of freedom %of a given system.

%Extracting the subleading contributions which are independent of the
%UV cutoff is an important task in the physical interpretation of the
%EE.

%Under a relevant deformation for an (1+1)-dimensional conformal field
%theory (CFT), the coefficient of the logarithmic UV divergent term in
%the EE is positive, monotonically decreasing, and stationary at fix
%points

As for the actual computation of the EE itself, it is usually known
that it is hard to evaluate the EE when the theory of interest is an
interacting one.  Actually, the majority of the computations of EE has
been done in free field theories.  However, the situation changes if a
field theory has its gravity dual in the context of AdS/CFT
correspondence.  According to the suggestion of \cite{Ryu:2006bv,Ryu:2006ef}
known as the
holographic EE (HEE),\footnote{For comprehensive review on the subject
  of HEE including related references, see \cite{Nishioka:2009un}.}  the EE of the
boundary field theory is given by the minimal surface area in the bulk
under the condition that the boundary of the minimal surface is the
entangling surface in the boundary theory.  Because the HEE concerns
only the geometric object, the minimal surface, which is simpler than
the direct quantum computation of boundary theory, it can be regarded
as a practical way to compute even the EE of interacting field theory,
at least in the case where the system is in its ground state.

Among many possible boundary field theories appearing in the AdS/CFT
correspondence, those originating from the explicit brane
configurations are particularly interesting because they are related
to the dual gravity or M/string theory stringently and believed to
play any important role in uncovering the nature of AdS/CFT
correspondence.  One such theory is
the $(2+1)$-dimensional $\mathcal{N}=6$
superconformal Chern-Simons matter theory with the gauge group
$U(N)_k\times U(N)_{-k}$ at Chern-Simons level $k$.  It describes
the $N$ M2-branes probing a $\mathbf{C}^4/\mathbf{Z}_k$ orbifold
and is called the Aharony-Bergman-Jafferis-Maldacena (ABJM)
theory~\cite{Aharony:2008ug}. One feature of this theory is that it
allows the supersymmetry preserving
mass-deformation~\cite{Hosomichi:2008jb,Gomis:2008vc}.
It has been shown in \cite{Kim:2010mr} that the gravity dual of the supersymmetric vacua of this mass-deformed ABJM (mABJM) theory for a
given $N$ and $k$ is identified with the half-BPS
Lin-Lunin-Maldacena (LLM) geometry~\cite{Bena:2004jw,Lin:2004nb}
with SO(2,1)$\times$SO(4)$\times$SO(4) isometry
in 11-dimensional supergravity. Interestingly, it was conjectured in
\cite{Lin:2004nb} that this type of LLM geometry (corresponding to $k=1$ case)
is dual to the supersymmetry preserving mass-deformation of the ${\cal N}=8$ CFT,
even before the appearance of the ABJM theory.

Since the mABJM theory is a deformation from the conformal ABJM theory,
it gives us a chance to study the behviour of the ABJM theory away from the
UV conformal fixed point with respect to the change
of the deformation parameter. At this point, the EE can be regarded as a good
measure for exploring such behaviour.
However, since the mABJM theory is highly interacting one, it is practically
too hard to compute its EE.  Fortunately, the dual geometries corresponding to
various supersymmetric vacua have been constructed~\cite{Cheon:2011gv} as
alluded to above and thus the HEE can be considered instead of EE.

In this paper, we are interested in the mABJM theory near the UV fixed point.
Our main goal is to compute the HEE's for general supersymmetric vacua and
to investigate the effect of the mass deformation from the viewpoint of
renormalization group (RG) flow.  The RG flow itself is derived from the
holographic renormalized EE (REE), which
has been proposed by Liu and Mezei \cite{Liu:2012eea} to define a UV finite
quantity from a given EE. It was shown  that the REE for any (2+1)-dimensional Lorentz invariant field theories  always monotonically decreases along the RG
trajectory~\cite{Casini:2012ei}. See also \cite{Klebanov:2012yf,Klebanov:2012va,Ishihara:2012jg,Nishioka:2014kpa,Kim:2014yca} for related works.
Especially in \cite{Kim:2014yca},  the present authors have done a study on the topic related with the REE of the mABJM theory, which is a preliminary work of our present work.
There, a circle was taken as the entangling surface and the HEE for
the most symmetric LLM geometry was calculated.   In the present work, we
extend the previous one.  We study the REE's of a strip as well as circular shaped
entangling surface on the general LLM geometries, which correspond to all possible
supersymmetric vacua of the mABJM theory.
We also discuss the validity of our computation in terms of gauge/gravity duality
in the large $N$ limit.

The organization of this paper is as follows.  In the next section, we
briefly review the supersymmetric vacuum structure of mABJM theory and the
corresponding dual LLM geometry in terms of droplet picture.
The HEE of the mABJM theory is studied in Sec.~\ref{hee}.  As mentioned
above, two types of entangling surface, strip and circular one,
are considered.  The configuration of droplet we take is quite
general except that it represents the weakly curved LLM geometry.
Based on the results of HEE, we compute the REE for each entangling
surface.
Finally, the summary of our results and discussion follow in
Sec.~\ref{secsum}.
In Appendix A, we find the relation between the mass parameter and the correlation
length which comes from a cutting off the tip of the minimal surface without mass-deformation. In Appendix B, we discuss the large $N$ behavior of the Ricci scalar
at the $y=0$ region.

%%%%%%%%%%%%%%%%%%%%%%%%%%%%%%
\section{Vacua of the mABJM theory and the LLM geometry}
%%%%%%%%%%%%%%%%%%%%%%%%%%%%%%

The ${\cal N}=6$ ABJM theory allows the supersymmetry preserving mass
deformation~\cite{Hosomichi:2008jb,Gomis:2008vc}
by imposing the same mass to 4-complex scalars and their superpartners. One intriguing
feature of the mABJM theory is that it has discrete Higgs vacua which are classified
by the partition of $N$. Here $N$ is the number of M2-branes in the ABJM theory.
The supersymmetric vacua~\cite{Kim:2010mr} of the mABJM theory
with Chern-Simons level $k=1$ have one-to-one correspondence with
the Lin-Lunin-Maldacena (LLM) background with SO(2,1)$\times$SO(4)$\times$SO(4) isometry in 11-dimensional supergravity~\cite{Bena:2004jw,Lin:2004nb}.
In this section we briefly review this correspondence and discuss the asymptotic properties of the LLM geometry.

\subsection{Supersymmetric vacua of the mABJM theory}

In this subsection, we summarize the supersymmetric
vacua~\cite{Kim:2010mr}
of the mABJM theory.
Before discussing it, we consider the classical vacuum equations,
which are obtained by setting the bosonic potential of the mABJM theory to zero~\cite{Aharony:2008ug}.
Since the SU(4) global symmetry of the original ABJM theory is broken to
SU(2)$\times$SU(2)$\times$U(1) symmetry in the mABJM theory, it is convenient to
split the SU(4)-symmetric 4-complex scalars into two SU(2)-symmetric complex scalars, i.e.,
\begin{align}
Y^A = (Z^a,\, W^{\dagger a}), \qquad
Y_A^\dagger = (Z_a^\dagger, \, W^a),
\end{align}
where $A=1,2,3,4$, $a=1,2$, and $Y_A^\dagger$ is the Hermitian conjugation of $Y^A$.
Then the vacuum equations are written as,
\begin{align}\label{Dtermeq}
&Z^aZ^\dagger_bZ^b-Z^bZ^\dagger_bZ^a=-\frac{\mu k}{2\pi} Z^a,\qquad W^{\dagger a}W_b W^{\dagger b}
-W^{\dagger b} W_b W^{\dagger a}=\frac{\mu k}{2\pi} W^{\dagger a},
\nn \\
&W_aZ^bW_b-W_bZ^bW_a=0,\qquad Z^bW_bZ^a-Z^aW_bZ^b=0.
\end{align}
Solutions of these equations in \eqref{Dtermeq} have been found in the form of the
GRVV matrices~\cite{Gomis:2008vc}.
Each vacuum solution is well represented as a direct sum of irreducible
rectangular $n\times (n+1)$ matrices, ${\cal M}_a^{(n)}$ $(a=1,2)$,
and their Hermitian conjugates,
$\bar {\cal M}_a^{(n)}$,\cite{Kim:2010mr,Cheon:2011gv}
\begin{align}\label{mat-1}
{\cal M}_1^{(n)}&=\left(\begin{array}{cccccc}
\sqrt{n\!}\!\!\!&0&&&&\\&\!\sqrt{n\!-\!1} \!\!&\!0&&&\\
&&\ddots&\ddots&&\\&&&\sqrt{2}&0&\\&&&&1&0\end{array}\right),
\qquad
{\cal M}_2^{(n)}&=
\left(\begin{array}{cccccc}0&1&&&&\\&0&\sqrt{2}&&&\\ &&\ddots&\ddots&&\\
&&&0\!&\!\!\sqrt{n\!-\!1}\!&\\&&&&0&\!\!\!\sqrt{n\!}\end{array}
\right).
\end{align}
In terms of these matrices, the vacuum solutions are
\begin{align}\label{ZW-vacua}
Z^a&=\sqrt{\frac{\mu k}{2\pi}}\left(\begin{array}{c}
\begin{array}{cccccc}\mathcal{M}_a^{(n_1)}\!\!&&&&&\\&\!\!\ddots\!&&&&\\
&&\!\!\mathcal{M}_a^{(n_i)}&&& \\ &&& {\bf 0}_{(n_{i+1}+1)\times n_{i+1}}
&&\\&&&&\ddots&\\&&&&&{\bf 0}_{(n_f+1)\times n_f}\end{array}\\
\end{array}\right),\nonumber
\end{align}
\begin{align}
W^{\dagger a}&=\sqrt{\frac{\mu k}{2\pi}}\left(\begin{array}{c}
\begin{array}{cccccc}{\bf 0}_{n_1\times (n_1+1)}&&&&&\\&\ddots&&&&\\
&&{\bf 0}_{n_i\times(n_i + 1)} &&&\\
&&& \bar{\mathcal M}_a^{(n_{i+1})}\!\!&&\\&&&&\!\!\ddots\!&\\
&&&&&\!\!\bar{\mathcal M}_a^{(n_f)}\end{array}\\
\end{array}\right),
\end{align}
where ${\bf 0}_{i\times j}$ denotes $i\times j$ null matrix.
Since $Z^a$ and $W^{\dagger a}$ are $N\times N$ matrices for the gauge group
${\rm U}(N)\times {\rm U}(N)$,\footnote{It can be also be extended to the vacuum solution
of the mass-deformed ABJ theory~\cite{Aharony:2008gk}
with the gauge group ${\rm U}(N)\times {\rm U}(N+l)$ having integer $l$.
See for the details~\cite{Cheon:2011gv}. In this paper, we mainly focus on the mABJM theory
with ${\rm U}(N)\times {\rm U}(N)$ gauge symmetry. } we have the following
constraints,
\begin{align}\label{levelmatch}
\sum_{n=0}^{N-1}\big[nN_n+(n+1)N_n'\big]=N,\qquad
\sum_{n=0}^{N-1}\big[(n+1)N_n+nN_n'\big]=N,
\end{align}
where $N_n$ $(N_n')$ denotes the number of blocks
of ${\cal M}_a^{(n)}$ $(\bar {\cal M}_a^{(n)})$.
%In this notation $N_0$ is the number of empty columns and
%$N_0'$ is that of empty rows.

Any combination of $(N_n,\, N_n')$ satisfying the constraint \eqref{levelmatch}
can be the solution of the vacuum equation \eqref{Dtermeq}.
However, it was found that the possible combinations of $(N_n, N_n')$
are much more than the number of the expected configurations~\cite{Gomis:2008vc}
in dual gravity theory, which are known as the LLM geometries. This problem
was resolved by introducing quantum fluctuations to classical vacua.
It was found that the occupation numbers for the quantum-level supersymmetric
vacua are further constrained by the Chern-Simons level $k$,
\begin{align}\label{susyvacuum}
0\le N_n\le k,\qquad 0\le N_n'\le k,
\end{align}
for every $n$~\cite{Kim:2010mr}.
Thus, only a subset of classical vacua remains supersymmetric at the quantum level.

\subsection{LLM geometry with $\mathbb{Z}_k$ quotient}

It was already conjectured in \cite{Lin:2004nb} that the LLM geometry with
SO(2,1)$\times$SO(4)$\times$SO(4) isometry in 11-dimensional supergravity should
correspond to the
${\cal N}=8$ effective field theory of M2-branes.
Subsequently, there has been much progress in this direction, for instance,
explicit matrix representation of discrete vacua~\cite{Gomis:2008vc},
supersymmetric vacua~\cite{Kim:2010mr}, one-to-one mapping between the supersymmetric
vacua of the mABJM theory and the LLM geometries for general $k$ and $N$~\cite{Cheon:2011gv}, etc..
See also \cite{Kim:2009ny,Hyun:2013sf} for other developments.

The LLM geometry with ${\mathbb Z}_k$ quotient is given by
\begin{align}\label{LLMgeom2}
ds^2 &= -G_{tt} \left(-dt^2 + dw_1^2 + dw_2^2\right) +G_{xx}(dx^2+dy^2)
+G_{\theta\theta} ds_{S^3/\mathbb{Z}_k}^2 + G_{\tilde\theta\tilde\theta}
ds_{\tilde S^3/\mathbb{Z}_k}^2
\end{align}
with
\begin{align}
ds^2_{S^3/\mathbb{Z}_k} &= d\theta^2 + \sin^22\theta \,d\phi^2
+\big((d\lambda + d\varphi/k) + \cos2\theta d\phi\big)^2,
\nn \\ \nn
ds^2_{\tilde S^3/\mathbb{Z}_k} &=d\tilde \theta^2 + \sin^22\tilde\theta
\,d\tilde\phi^2
+\big((-d\lambda + d\varphi/k) + \cos2\tilde \theta
d\tilde\phi\big)^2,
\end{align}
where
\begin{align}
-G_{tt} & = \left(\frac{4\mu_0^2 y\sqrt{\frac14 - z^2}}{f^2}\right)^{2/3},
\nn \\
G_{xx} & = \left(\frac{f \sqrt{\frac14-z^2}}{2\mu_0 y^2}  \right)^{2/3},
\nn \\
G_{\theta\theta} & = \left(\frac{f y \sqrt{\frac12 + z}}{2\mu_0 \left(\frac12 - z\right)}   \right)^{2/3},
\nn \\
G_{\tilde\theta\tilde\theta} & = \left(\frac{f y \sqrt{\frac12 - z}}{2\mu_0 \left(\frac12 + z\right)}   \right)^{2/3} , \nn \\
f(x,y)  & = \sqrt{1- 4z^2 - 4 y^2 V^2} .
\end{align}
The mass parameter $\mu_0$ corresponds to turning on a non vanishing
4-form field strength.
The LLM geometry in \eqref{LLMgeom2} is completely determined
in terms of $z(x,y)$ and $V(x,y)$,
\begin{align}\label{zV}
z(x,y) &= \sum_{i=1}^{2N_B +1}\frac{(-1)^{i+1}(x-x_i)}{2\sqrt{(x-x_i)^2 + y^2}},
\nn \\
V(x,y) &= \sum_{i=1}^{2N_B +1}\frac{(-1)^{i+1}}{2\sqrt{(x-x_i)^2 + y^2}},
\end{align}
where $N_B$ is the number of the black droplets and
$x_i$'s represent the locations of the boundary lines between the black
and white strips.
The black and white strips in such a droplet representation indicate the $\mp \frac12$
values of the function $z$ along the $y=0$ boundary.
For the detailed prescription of the droplet representation for general $k$, see \cite{Cheon:2011gv}.
%\subsubsection{Asymptotic expansion}

\section{HEE of the mABJM Theory}
\label{hee}

%==================================================
\begin{figure}
\begin{center}
\includegraphics[width=9cm,clip]{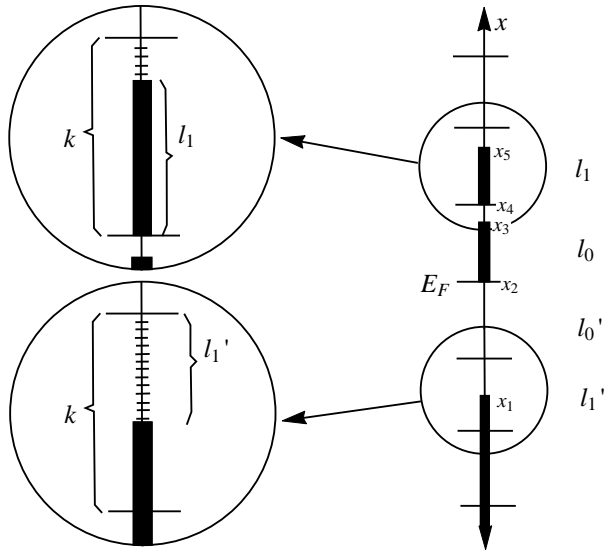}
\end{center}
\caption{
\label{fig:droplet}
An example for $N_B=2$: $E_F$ is the Fermi-energy which is the level of black droplet defined when all excited black droplets sink down. The $k$ unit length divides the $x$-axis into sections denoted by indices $n=0,1,2,\ldots$, and each section has $l_n$ or $l_n'$ which is the length of black part or white part, respectively. They are identified with $N_n$ and $N_n'$ describing a field theory vacua.   }
\end{figure}
%==================================================

%%==================================================
%\begin{figure}
%\begin{center}
%\includegraphics[width=9cm,clip]{droplet01.eps}
%\end{center}
%\caption{
%\label{fig:droplet}
%Example for $N_B=3$:$E_F$ is the Fermi-energy which is the level of black droplet defined %when all excited black droplets sink to the ground state. $k$ unit lengths define $l$ or $l'$ %sections and each section has the number $N_l$ or $N_{l'}$, which is the length of black %droplet or white droplet in term of the unit length. In addition, $x_i$'s denote boundaries of %black and white droplets.   }
%\end{figure}
%==================================================

The LLM geometry introduced in the previous section is
asymptotically AdS$_4\times {\rm S}^7/{\mathbb Z}_k$, which means that
the conformal symmetry is restored in the UV limit
and the dual field theory becomes the ABJM theory without mass deformation. Due to the mass deformation, the conformal symmetry of the system is broken and the dual geometry
should be modified in the deep IR region. This correspondence makes it possible to investigate the effect of the mass deformation on the HEE near the UV conformal fixed point of the ABJM theory.
Interestingly, it was shown that the REE
derived from the HEE at the UV fixed point is consistent with the free energy
of the ABJM theory obtained by the localization technique on
${\rm S}^3$~\cite{Kapustin:2009kz}.
In this section, we compute the HEE and investigate the REE near the UV fixed point of the mABJM theory with a small mass deformation for two types of entangling surface, strip and disk.
For consistency check of our results, we will discuss the validity of the dual LLM geometry .

\subsection{Strip}

First, let us consider the HEE of the strip defined at the boundary of the LLM geometry.
Unlike the case of AdS$_5 \times S^5$,
where the role of the compact manifold is trivial, the LLM geometry is not a simple product space
so that one should be careful in evaluating the HEE. For the
HEE of the strip,  we regard a $9$-dimensional surface embedded in the LLM geometry
which is called a holographic entangling surface for simplicity.
Its boundary of course is identified with the boundary of the strip.
If the coordinates of the holographic entangling surface are denoted by $\sigma^i$ with $i=1, \cdots, 9$, the induced metric can be represented as a functional of the embedding function $X^M (\sigma^i)$
\begin{align}\label{indmet}
g_{ij} = G_{MN}  \ \frac{\partial X^M\partial X^N}{\partial \sigma^i\partial\sigma^j} ,
\end{align}
where $G_{MN}$ is the $11$-dimensional LLM metric. Then, the surface shape is governed
by the following action
\begin{align}\label{gammaA0-1}
\gamma_A = \int d^9\sigma\, \sqrt{\det g_{ij}}.
\end{align}
It was conjecture in \cite{Ryu:2006bv,Ryu:2006ef} that the minimal area corresponding to the on-shell action
is proportional to the HEE of the strip
\begin{align}
S_A = \frac{{\rm Min}(\gamma_A)}{4G_N}.
\end{align}
The boundary space of the LLM geometry can be represented by $R^{1,2} \times S^7/\mathbb{Z}_k$.
If we consider a static strip configuration, it should be extended in the $2$-dimensional
noncompact flat space and wrap the $7$-dimensional compact manifold.
Let us suppose that the strip is extended in $w_1$-direction infinitely and has a finite width $l$ in  $w_2$-direction. Then, the holographic entangling surface describing the HEE of the strip can be
parameterized as follows
\begin{align}
&w_1=\sigma^1,  \quad \left(-\frac{L}{2}\le w_1 \le \frac{L}{2}\right),
\nn \\
&w_2 = \sigma^2,  \quad \left(-\frac{l}{2}\le w_2 \le \frac{l}{2}\right),
\nn \\
&\alpha = \sigma^3,  \,
\theta=\sigma^4,\, \phi=\sigma^5,\,
\tilde\theta=\sigma^6,\, \tilde\phi= \sigma^7,\, \lambda = \sigma^8,~ \varphi = \sigma^9  ,
\end{align}
where the infinite length of $w_1$ is regularized to $L$ for convenience.
The holographic entangling surface is also extended in the radial direction $r$ which generally
becomes a function of the noncompact coordinates. However, the translation symmetry
in the $w_1$ direction requires $r$ to be a function of $w_2$ only, $r=r(w_2)$.

Substituting the LLM metric into the induced metric formula leads to
\begin{align}
ds^2 =
%|G_{tt}|((d w_1)^2 + (d w_2)^2  ) +   G_{xx}\left( r'^2 (d w_2)^2 + r^2 d\alpha^2   \right) +  %G_{\theta\theta}ds^2_{S^3/\mathbb{Z}_k} +  G_{\tilde \theta \tilde \theta }ds^2_{\tilde %S^3/\mathbb{Z}_k}  \\=
|G_{tt}|(d w_1)^2 + ( |G_{tt}| + G_{xx}r'^2  )(d w_2)^2   +   G_{xx}  r^2 d\alpha^2    +  G_{\theta\theta}ds^2_{S^3/\mathbb{Z}_k} +  G_{\tilde \theta \tilde \theta }ds^2_{\tilde S^3/\mathbb{Z}_k} .
\end{align}
Then the action of the holographic entangling surface, after performing
the integrations over all angles but $\alpha$, reduces to
\begin{align}
\gamma_A %&=\int_{-L/2}^{L/2} dw_1  \int_{-l/2}^{l/2} dw_2 \int_0^{\pi} d\alpha
%\int_0^{\pi/2 } d\theta \int_0^{ \pi} d\phi  \int_0^{\pi} d\lambda
%\int_0^{\pi/2 } d\tilde \theta \int_0^{ \pi} d\tilde\phi  \int_0^{2\pi} d\varphi
% \sqrt{\det g_{ij}}
%\nn \\ &
=\frac{4\pi^4 L}{k} \int_{-l/2}^{l/2} dw_2 \int_0^{\pi} d\alpha~ r\sqrt{\left|G_{tt}\right|G_{xx} G_{\tilde\theta \tilde\theta}{}^3 G_{\theta \theta }{}^3 \left(\left|G_{tt}\right|+G_{xx} r'^2\right)}  \  ,
\end{align}
where  the prime means a derivative with respect to $w_2$.
Note that $x$, $y$ and $r$ have length square dimension.
Let us introduce two dimensionless
variables and a new radial coordinate with length dimension
\begin{align}
\tilde x = \frac{4 x}{R^2} \quad , \quad \tilde y = \frac{4 y}{R^2} \quad
{\rm and} \quad u = \frac{R^3}{4r} ,
\end{align}
which are related to each other as follows:
\begin{align}
\tilde x = \frac{R}{u} \cos \alpha \quad {\rm and} \quad \tilde y = \frac{R}{u} \sin \alpha \ .
\end{align}
By using the relation between parameters
\begin{align}\label{gammaA3}
R= (32\pi^2 k \tilde N)^{1/6} l_{\rm P}  ,
%{\rm and} \quad p = \frac{R^3}{2\sqrt{2}}\, \mu_0 \ ,
\end{align}
the action can be rewritten as
\begin{align}\label{gammaA5}
\gamma_A &=\frac{\pi^4 L R^9}{32 k\mu_0} \int_{-l/2}^{l/2} dw_2 \int_0^{\pi} d\alpha \,
\frac{f \ \sin^2\alpha }{u^3} \sqrt{1+ \frac{f^2 u'^2}{4\mu_0^2 \sin^2\alpha\,
u^2}}  \ ,
\end{align}
with
\begin{align}
f & = \sqrt{1- 4  z^2 - 4  y^2  V^2} \,, \nn \\
 z  &= \sum_{i=1}^{2 N_B + 1}\frac{(-1)^{i+1}( x- x_i)}{2\sqrt{( x- x_i)^2 +  y^2}} \notag \\
 & = \frac{1}{2}\left[ \cos\alpha + \sum_{k=1}^{\infty}\sum_{i=1}^{2N_B +1} \left(  \frac{x_i}{r} \right)^k (-1)^{i+1}\left( P_1(\cos \alpha)P_k(\cos \alpha) - P_{k-1}(\cos \alpha)  \right)  \right] \,,
\nn \\
 V  &= \sum_{i=1}^{2N_B+1}\frac{(-1)^{i+1}}{2\sqrt{( x- x_i)^2 +  y^2}}
=\frac{1}{2} \sum_{k=0}^\infty \sum_{i=1}^{2N_B+1} (\frac{1}{r})^{k+1}  x_i^k (-1)^{i+1} P_k (\cos\alpha) \,, \label{fzv}
\end{align}
%\subsubsection{Small mass expansion}
where $N_B$ is the number of black droplets.

Generally speaking, the Lagrangian density for the minimal surface is very complicated in the full LLM geometry and so it is hard to avoid difficulty in performing angle integration. However, our main goal is to see the small mass deformation effect. For this, it is enough to take into account the limit $\mu_0 \ll r$ rather than the full geometry. In this approximation the mass deformation effect on the HEE appears as deviation from the HEE obtained in AdS$_4$. More specifically, using (\ref{fzv}),
the function $f$ in the small mass limit is expanded as follows.
\begin{eqnarray} \label{rel:f}
f = D_0 \, \mu _0 \, u \sin \alpha  \left[1 + D_1 \, \mu _0  \, u \cos \alpha   +
( D_2 +D_3 \,  \cos (2 \alpha ))  \, \mu _0^2 \, u^2  + {\cal O} (\mu_0^3)\frac{}{} \right] ,
\end{eqnarray}
where
\begin{align}
&D_0 = \sqrt{2} \sqrt{C_{2}-C_{1}^2} ~,~D_1= -\frac{ \left(C_{1} C_{2}-C_{3}\right)}{\sqrt{2}}~, \notag \\
&D_2= \frac{1}{16}  \left(-5 C_{2}^2-2
  \left(C_{3}-C_{1} C_{2}\right){}^2-4 C_{1} C_{3}+9 C_{4}\right) ~ ,
  \notag \\
&D_3= \frac{1}{16} \left(-3 C_{2}^2-2 \left(C_{3}-C_{1} C_{2}\right){}^2-12 C_{1} C_{3}+15 C_{4}\right)~  .
\label{coD}
\end{align}
The coefficient $C_{k}$ appearing in the above formula is defined as
\begin{align}
C_{k} = \sum _{i=1}^{2N_b + 1} (-1)^{i+1}    \left( \frac{ \hat x_i}{\sqrt{N k}}  \right)^k ,
\end{align}
and satisfies $ \left( C_{2}-C_{1}^2 \right) =2$. Here $\hat x_i$ is defined as
$ x_i = 2\pi l_p^3  \mu_0 \hat x_i$.
When the Chern-Simon level $k$ is equal to 1, a vacuum of the mass deformed ABJM theory can be represented by a Young diagram with $\hat x_i$, whose area is given by $N$. This area implies the number of M2-branes \cite{Lin:2004nb}.

%%%%%%%%%%%%%%%%%%%%%%%%%%%%%%%%%%%%%%%%%%%%%%%%%%%%%%%%%%%%%%%
\begin{figure}%[!h]
\centerline{\epsfig{figure=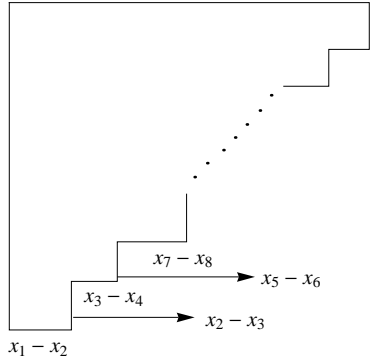,height=60mm}}
\caption{
\small k=1 case : The area of the Young-diagram is given by $\left( 2\pi\mu_0 l_p^3  \right)^2 N$.   }
\label{Fig1}
\end{figure}
%%%%%%%%%%%%%%%%%%%%%%%%%%%%%%%%%%%%%%%%%%%%%%%%%%%%%%%%%%%%%%%

Expanding the action with the small $\mu_0$, one can easily perform integration for the angle $\alpha$. Up to
$\mu_0^2$ order, $\gamma_A$ is expanded as
\begin{align}
\gamma_A \approx  & \frac{\pi ^4 L R^9 }{12 k} \int_{-l/2}^{l/2} d w_2
\left( \frac{ \sqrt{u'^2+1}}{u^2} \right. \nn \\
& \left. + \frac{ \mu _0^2  \left(2 \left(D_1^2+10 D_2-6 D_3\right) u'^4+3 \left(D_1^2+10 D_2-6 D_3\right) u'^2+10 D_2-6 D_3\right)}{10  \left(u'^2+1\right)^{3/2}} \right) .
\end{align}
If we regard $w_2$ as time, then $\gamma_A$ can viewed as an action of the mechanical problem.
Since the Lagrangian is independent on $w_2$, one may construct a conserved Hamiltonian as follows.
\begin{align}
H= -\frac{\pi ^4 L R^9}{12 k u^2 \sqrt{u'^2+1}}-\frac{\pi ^4 \mu _0^2 L R^9 \left(\left(-3 D_1^2+10 D_2-6 D_3\right) u'^2+10 D_2-6 D_3\right)}{120 k \left(u'^2+1\right)^{5/2}}~~.
\end{align}
At the turning point denoted by $u_0$, $u'(w_2)$ vanishes. Applying this condition, the Hamiltonian turns out to be
\begin{align}
H = -\frac{\pi ^4 L R^9}{12 k u_0^2}  +
\frac{\pi ^4 \left(6 D_3-10 D_2\right) \mu _0^2 L R^9}{120 k} .
\end{align}
Comparing above two expressions, $u'$ can be written in terms of $u$
\begin{align}
u' = - \frac{\sqrt{u_0^4-u^4}}{u^2} + \frac{\mu _0^2 \left(-3 D_1^2 u^{10}+\left(3 D_1^2-10 D_2+6 D_3\right) u_0^4 u^6+2 \left(5 D_2-3 D_3\right) u_0^{10}\right)}{10 u^2 u_0^4 \sqrt{u_0^4-u^4}}
\end{align}
By integration of this equation,
the width $l$ and the minimal area $\gamma_A$ are presented as functions of $u_0$ up to $\mu_0^2$ order,
\begin{align}
l = & \sqrt{\frac{2}{\pi }} u_0 \Gamma \left(\frac{3}{4}\right)^2
\notag \\
&+ \frac{\mu _0^2 u_0 \left(5 \left(3 D_1^2+35 D_2-21 D_3\right) u_0^2 \Gamma \left(\frac{1}{4}\right) \Gamma \left(\frac{5}{4}\right)+21 \left(3 D_3-5 D_2\right) u_0^2 \Gamma \left(\frac{3}{4}\right)^2\right)}{105 \sqrt{2 \pi }} , \nn \\
\gamma _A = &\frac{\pi ^4 L\text{  }R^9}{6 k}\text{  }\left(\frac{1}{\epsilon } -\frac{\Gamma \left(\frac{3}{4}\right)^2}{\sqrt{2 \pi } u_0}+\frac{\mu _0^2 u_0 \left(15 D_1^2 \Gamma \left(\frac{1}{4}\right)^2+7 \left(5 D_2-3 D_3\right) \left(5 \Gamma \left(\frac{1}{4}\right)^2-4 \Gamma \left(\frac{3}{4}\right)^2\right)\right)}{280 \sqrt{2 \pi }}\right) , \label{preEE}
\end{align}
where $\epsilon$ denotes a UV cutoff. Substituting $u_0$ into $\gamma_A$, the strip entanglement entropy up to $\mu_0^2$ order reads in terms of $l$
\begin{align}
S_A = \frac{\gamma_A}{4 G_N}=\frac{\pi ^4 L\text{  }R^9}{24 G_N k}
\left( \frac{1}{\epsilon }  -\frac{\Gamma \left(\frac{3}{4}\right)^4}{\pi  l} - \mu _0^2 l   \frac{\left(21 D_3-3 D_1^2-35 D_2\right) \Gamma \left(\frac{1}{4}\right)^4}{336 \pi ^2}  \right) ,
\label{stripSA}
\end{align}
where $G_N= (2\pi l_{{\rm P}})^9/(32\pi^2)$ denotes a $11$-dimensional Newton's constant with the Planck length $l_{{\rm P}}$.
The first and second terms on the right had side
are consistent with the HEE obtained in AdS$_4$, as mentioned before,
and the third term is the leading correction caused by the mass deformation in the small mass limit.
According to \cite{Klebanov:2012va}, we can define a holographic $c$-function of the strip
\begin{align}
{\cal F}_{{\rm strip}}(l) & \equiv l^2 \partial_l \hat S_A \nn \\
& =  \frac{\pi ^3 R^9 }{24 G_N k} \left[ \Gamma \left(\frac{3}{4}\right)^4
-\mu _0^2  l^2   \left( \frac{\pi \, \Gamma \left(\frac{1}{4}\right)^2 \left(21 D_3 - 3 D_1^2-35 D_2\right)   }{168  \Gamma \left(\frac{3}{4}\right)^2}   \right)\right] ,
\label{Fstrip}
\end{align}
where $\hat S_A \equiv  S_A/L$.
If the coefficient of the $\mu_0^2 l^2$ is negative,  ${\cal F}'_{{\rm strip}}(l)$ becomes negative, which implies that
the holographic $c$-function monotonically decreases along the RG flow.

For more concrete example, now let us take into account the symmetric configuration where the parameters are
given by
\begin{align}
D_1=0, \quad  D_2=-1/8, \quad {\rm and} \quad  D_3=9/8. \label{sym config}
\end{align}
Then the free energy or $c$-function of the symmetric strip reduces to
\begin{align}
{\cal F}_{{\rm strip}}(l) = \frac{\pi ^3 R^9 \Gamma \left(\frac{3}{4}\right)^4}{24 k G_N}- \mu_0^2 l^2 \frac{ R^9   \pi ^2 \Gamma \left(\frac{1}{4}\right)^4 }{288 k G_N} .
\end{align}
In this case the coefficient of the $\mu_0^2 l^2$ is a negative number. This fact implies
that the holographic $c$-function shows the monotonically decreasing behavior along the RG flow.

Before going to the disk case, we would like to give a comment on another way of the mass deformation. In \cite{Ryu:2006ef}, the authors considered the mass deformation of CFT in a bottom-up approach. So it is meaningful to make a comparison between our top-down result and theirs. We discuss the identification between them in Appendix A.

%%%%%%%%%%%%%%%%%%%%%%%%%%%%%
\subsection{Disk}
%%%%%%%%%%%%%%%%%%%%%%%%%%%%%

We now turn to the REE of a disk near the UV fixed point.
Let us take a circular region with radius $l$ on the two
spatial directions of the boundary noncompact manifold.
The $9$-dimensional holographic entangling surface with two noncompact directions is embedded
into the target space \eqref{LLMgeom2} as
\begin{align}\label{mapping1}
&u= u(\rho),  \, w_1 = \rho \cos \sigma^1, \, w_2 = \rho \sin \sigma^1,
\,
\nn \\
&\alpha = \sigma^3, \, \theta=\sigma^4,\, \phi=\sigma^5,\,
\tilde\theta=\sigma^6,\, \tilde\phi= \sigma^7,\, \lambda = \sigma^8,~ \varphi = \sigma^9 ,
\end{align}
where the radial coordinate of AdS$_4$ is given by $u = R/\sqrt{\tilde x^2 + \tilde y^2}$
and is a function only of $\rho$  due to the rotation symmetry in the $(w_1 , w_2)$ plane.
$\alpha=\tan^{-1}(\tilde y/\tilde x)$ is the angle in the $(x , y)$ plane and the range of $\rho$
is given by $ 0\le  \rh \le  l$.
The action describing the holographic entangling surface, after integrating  out angular variables
of the compact space,
reduces to
\begin{align}\label{gammaA}
\gamma_A
=\frac{\pi^5 R^9}{16 k \mu_0}\int_0^l d\rho \,\int_0^{\pi} d\alpha \,
\frac{f \rho  \sin ^2 \alpha}{u^3}
\sqrt{1+ \frac{f^2  u'^2}{4 \mu_0^2 \sin^2 \alpha   u^2}},
\end{align}
where the prime means a derivative with respect to $\rho$, and $f$ in the small mass
limit is given  in (\ref{rel:f}).  In this small mass limit, the $\alpha$ integration
up to $\mu_0^2$ order leads to
\begin{align}\label{act:deformed}
&\gamma_A = \frac{\pi^5 R^9}{6 k} \int_0^l d\rho  \,
\rho\, \bigg[\frac{\sqrt{1 + u'^2}}{u^2}   + \frac{(5 D_2 -3 D_3 ) \mu_0^2}{5\left( 1 + u'^2  \right)^{3/2}}
+ \frac{   (D_1^2 +10 D_2 - 6 D_3)(3 u'^2 + 2 u'^4) \mu_0^2}{
10\left( 1 + u'^2  \right)^{3/2}} \bigg] ,
\end{align}
where the normalization used in the previous section $D_0=2$ is adjusted.
Note that, unlike the strip case, there is no conserved charge due to the explicit dependence on $\rho$.
So we can not apply the method used in the previous section to the disk case. Here we follow a different strategy.

The minimum value of $\gamma_A$ is given by the on-shell action.
In the $\mu_0\to 0$ limit, $\gamma_A$ should be reduced to that of the AdS$_4$ up to
an overall factor caused by the volume of the $7$-dimensional compact manifold.
In this zero mass limit, it is well-known that a circle appears as a special solution satisfying the boundary conditions, $u'(0)=0$
and $u(l)=0$,
\begin{align}
u_0(\rho) = \sqrt{l^2 - \rho^2} .
\end{align}
In order to figure out the mass deformation effect near the UV fixed point,
we can take into account a small mass perturbation around the known circular solution.
The leading contribution appears at $\mu_0^2$ order, so we take an ansatz
\begin{align}\label{pertu}
u (\rho) = u_0 (\rho)  + (\mu_0 \, l)^2  \ \delta u (\rho)    .
\end{align}
Then, the fluctuation field $\delta u$ is governed by an inhomogeneous second order differential equation
\begin{align}
0 = &\delta u'' +\frac{ \left(l^2-2 \rho ^2\right)}{l^2 \rho -\rho ^3} \ \delta u'
- \frac{2  l^2}{\left(l^2-\rho ^2\right)^2} \ \delta u \nn \\
&+ \frac{D_1^2 \left(-6 l^4+14 l^2 \rho ^2-9 \rho ^4\right)-2 \left(5 D_2-3 D_3\right) l^2 \left(3 l^2-2 \rho ^2\right)}{5 l^4 \sqrt{l^2-\rho ^2}} ,
\end{align}
which allows two integration constants. For the fluctuation solution to be determined unambiguously,
we must impose two natural boundary conditions.
In the asymptotic region ($u \to 0$), the effect of the mass deformation is negligible, so
the deformed solution should reduce to the undeformed one, $u(l) = u_0 (l)$.
This fact implies that the fluctuation field $\delta u (l)$ vanishs at the boundary.
Assuming that the action in (\ref{gammaA}) is regular, the holographic entangling surface should  be smooth.
This smoothness, together with the rotational symmetry in the $(w_1,w_2)$ plane, enforces $\delta u'(0)=0$
at the turning point. Imposing these two boundary conditions fixes the fluctuation field uniquely
\begin{align}\label{delta}
\delta u(\rho)  &= \frac{l^3}{300\sqrt{1- (\rho/l)^2}}\Big[
18 D_1^2 (\rho/l)^6 - (61 D_1^2 +100 D_2 -60 D_3) (\rho/l)^4 \nn \\
&+(112 D_1^2 + 700 D_2 - 420 D_3) (\rho/l)^2
- 69 D_1^2 - 600 D_2 + 360 D_3 \nn \\
&-8 (11D_1^2 + 125 D_2 - 75 D_3)
 \Big(\tanh^{-1}\sqrt{1- (\rho/l)^2}  + \ln (\rho/l)
-\sqrt{1- (\rho/l)^2} \Big)
\Big].
\end{align}
In Fig 2. we plot the
deformed holographic entangling surface in the symmetric case in which the mass deformation pushes
the turning point toward the AdS$_4$ center.

%==================================================
\begin{figure}
\begin{center}
\includegraphics[width=8cm,clip]{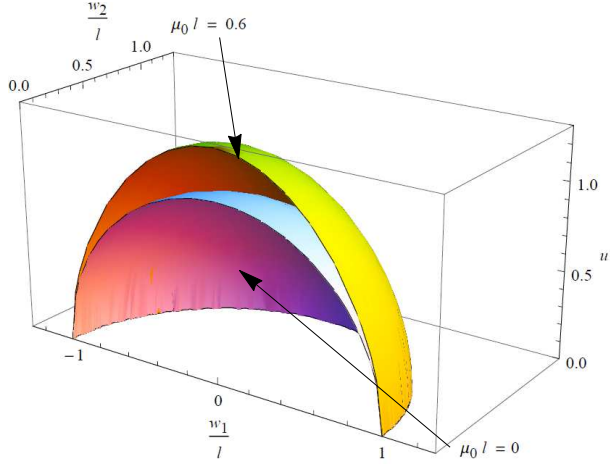}
\end{center}
\caption{
\label{fig:msurface}
 Minimal surfaces for the strip case : The upper surface is a deformed minimal surface due to the mass deformation and the lower surface is for the conformal case}
\end{figure}
%==================================================

After integrating over $\rho$, the on-shell action up to $\mu_0^2$ order leads to
the HEE of the disk in terms of the radius $l$
\begin{align}\label{Sdisk1}
S_{{\rm disk}} = \frac{\pi ^5  R^9}{24 k G_N}\left(\frac{l}{\epsilon} -1
-\frac{75 D_3-11 D_1^2 - 125 D_2  }{75}\,(l \mu_0)^2\right),
\end{align}
where the UV cutoff in the $u$ coordinate is denoted by $\epsilon$.
In the above, the first two terms on the right hand side correspond to the HEE of a disk in the
ABJM theory and the last is the first correction caused by the mass deformation.
For instance, in the simplest symmetric case of the droplet picture with $k=1$,
the parameters $x_i$
%representing the Coulomb branch
have
\begin{align}
x_1= -\sqrt{N} \ ,\quad  x_2=0 \ , \quad {\rm and}  \  x_3=\sqrt{N} ,
\end{align}
which give rise to $D_1=0$, $D_2=-1/8$ and $D_3=9/8$.
Then, the HEE of the symmetric configuration is given by
\begin{align}
S_{{\rm symm}} = \frac{\pi ^5  R^9}{24 k G_N}\left(\frac{l}{\epsilon} -1
-\frac{4}{3}\,(l \mu_0)^2\right) .
\end{align}
The free energy corresponding to the $c$-function of this system then reduces
to
\begin{align}
{\cal F}_{\rm  symm} \equiv & \left( l \frac{\partial}{\partial l}  - 1 \right) S_{\rm symm}\nn \\
= & \frac{\pi ^5 R^9 }{72 k G_N} \left[ 3 - 4 ( l \mu _0)^2 \frac{}{} \right]  .
\end{align}
This result shows that at a given $\mu_0$ the free energy
decreases along the RG flow when the system size $l$ increases. As expected,
this result coincides with the $F$-theorem, ${\cal F}_{\rm symm}' < 0$.

For general droplets, we finally obtain
the REE  up to $\mu_0^2$ order
\begin{align}\label{Fdisk3}
{\cal F}_{{\rm disk}}(l) = {\cal F}_{ABJM} -
\frac{\pi ^5  R^9(75 D_3-11 D_1^2 - 125 D_2 )}{1800 kG_N}\,(l \mu_0)^2,
\end{align}
where ${\cal F}_{ABJM} = \frac{\pi ^5  R^9}{24 k G_N}$ is the free energy of the original
ABJM theory.
The REE counts the effective degrees of freedom of a given system
at the length scale $l$ and is expected to play a role of a $c$-function in
the holographic point of view.
Since the mass deformation we consider is a relevant deformation
with the dimensionless coupling constant $g= l \mu_0$,
the monotonic decreasing of the REE along the RG flow is guaranteed by
the following relation
\begin{align}
75 D_3-11 D_1^2 - 125 D_2 > 0 .
\end{align}
If this relation is satisfied, our result supports the $F$-theorem in 3-dimensional field
theory~\cite{Jafferis:2011zi,Myers:2010tj}.
In general droplets, it seems to be difficult to prove the above inequality.
We first take into account the symmetric droplet configurations.
In these cases, the validity of the dual LLM geometry, as will be shown in the next subsection,
depends on the parameter regions. In the regions where the dual LLM geometry
is weakly curved, the above inequality is really satisfied.
As a result, the REE in \eqref{Fdisk3} shows the desired holographic
$c$-function behavior near the UV fixed point
and, as the system size $l$ increases, monotonically decreases consistently with the $F$-theorem along the RG flow.

\subsection{Validity}

The results of REE given in \eqref{Fstrip} and \eqref{Fdisk3} are
for general LLM geometries near the UV fixed point for the strip and the disk cases, respectively.
To guarantee their validity, we have to check whether
the LLM geometries we have considered are weakly curved everywhere in
the large $N$ limit.
%Especially we need to consider the area near $y=0$, where can be strongly
%curved even in the large $N$ limit~\cite{}.
Here we sketch the validity of our calculations of the REE in the point of
view of the gauge/gravity duality.
The validity of the LLM geometry for the symmetric droplet
case has already been considered in \cite{Hyun:2013sf}, where it was
shown that the magnitude of the curvature scalar
is decreasing from the droplet point $y=0$ as $y$ increases.
On the contrary, the validation is broken for
the cases where the curvature scalar does not decrease and
remains constant near $y=0$ in the large $N$ limit.
Therefore, in checking the validity, it is sufficient to investigate the behavior of the curvature
at the droplet point, $y=0$.  Following the same logic,
we extend the discussion to more general droplet cases.

Let us consider the general droplet characterized by the data $x_i$,
which specify $z(x,y)$ and $V(x,y)$ of \eqref{zV} describing the full
geometry.
As one can find in Appendix B, if a given geometry does not have
any strongly curved region, it is represented a droplet whose
associated Young-diagram has only the long edges
of order $\sqrt{N}$~\cite{Hyun:2013sf,Kim:2014yca}.
If there is a strongly curved region, the validity of
the HEE is not guaranteed. As an example we consider the droplet corresponding to
the rectangular shaped Young-diagram with sides of lengths $w$ and $b$.
By parametrizing the lengths as $w = \frac{\sqrt{N}}{\hat\sigma},\,\,
b= \hat\sigma\sqrt{N}$, (\ref{coD}) leads to
\begin{align}
D_1 = -\frac{\tilde\sigma}{\sqrt{2}}, \quad
D_2 = \frac18 \big(\tilde\sigma^2-1\big),\quad
D_3= \frac18 \big(5\tilde\sigma^2+9\big),
\end{align}
where $\tilde\sigma = \hat\sigma - \frac1{\hat\sigma}$.
Here $k=1$ is taken for simplicity.
Using these relations we obtain the REE for the strip and the disk,
\begin{align}\label{stripDisk}
{\cal F}_{{\rm strip}}(l) &=  \frac{\pi ^3 R^9 }{24 G_N } \left[ \Gamma \left(\frac{3}{4}\right)^4
-    \frac{\pi \, \Gamma \left(\frac{1}{4}\right)^2 \left(29\tilde\sigma^2 + 112\right)   }{672  \Gamma \left(\frac{3}{4}\right)^2}  \,(l \mu_0)^2\right] ,
\nn \\
{\cal F}_{{\rm disk}}(l) &= \frac{\pi^5R^9}{24 G_N}\left(1 -
\frac{103\tilde\sigma^2 + 400 }{300 }\,(l \mu_0)^2\right).
\end{align}

From (\ref{calR}), we see that the finite $\hat\sigma$ gives
weakly curved LLM geometry.
So we expect that our result given in \eqref{stripDisk}
is valid. However, in the case $\hat\sigma \sim \sqrt{N}$ or $\hat\sigma \sim \frac1{\sqrt{N}}$, the curvature scalar near $y=0$ region remains finite in the large $N$ limit. Therefore, the gauge/gravity duality may not be valid anymore.
In turn, the leading contributions of the mass-deformation
to the REE's of \eqref{stripDisk} are going to
diverge, and thus invalidate the gauge/gravity duality.
In conclusion, as we discussed previously, in order to have valid results of HEE we have to consider the Young-diagram including only long edges
of order $\sqrt{N}$.

%%%%%%%%%%%%%%%%
\section{Summary}
\label{secsum}
%%%%%%%%%%%%%%%%

Following the gauge/gravity duality, we have investigated the REE of the mass deformed ABJM theory and its RG flow.
To do so, we have taken into account the LLM geometry corresponding to vacua of the mABJM
theory which can be reinterpreted as droplets in the droplet picture.
In general, the REE crucially depends on the droplet configuration, so it is a formidable
take to find the analytic form of the general REE in the entire region.
In this paper, we focused on the UV region
where, due to the relatively small mass deformation,
the perturbative and analytic studies on the mass deformation effect are possible.
The entanglement entropy is an important concept to understand the degrees of freedom of a physical system. Interestingly, it was shown that the REE of the disk
is associated with the free energy of an odd dimensional quantum field theory.
The REE generally depends on the shape of the system we consider so that
different shaped-systems result in different REE's. Here,
two types of the REE with the strip and disk shapes have been regarded.

The  LLM geometry near the asymptotic boundary can be expanded in terms of the Legendre polynomials.
In this region, the REE's of the strip and disk are given by nontrivial functions of the expansion coefficients.
We have shown the explicit dependence of the mass deformation in those two shapes. The first correction of the REE appears at $ (l \mu_0)^2$ order, which implies that the variation
of the REE with respect to coupling $g = l \mu_0$ always vanishes as $l$ goes to zero.
Therefore, the REE at the UV fixed point is always stationary.\footnote{Stationarity near UV fixed point in (2+1)-dimensions was discussed in \cite{Klebanov:2012va,Nishioka:2014kpa}. Especially in \cite{Nishioka:2014kpa}, the author
classified the behavior of the REE according to the dimension the perturbed relevant operators.}
Near the UV fixed point, the variation of the REE  explains the nontrivial dependence
on the deformation parameter which is related to $c$-functions along the RG flow.

In a simple example with a rectangular shaped Young diagram, if
the ratio between width and height is given by $1$ in the large $N$ limit, it describes a symmetric
droplet configurations. In this case, the REE's of the strip and disk
have a negative slope. So the free energy corresponding the REE monotonically decreases
along the RG flow and satisfies the $F$-theorem.
In the asymmetric case slightly deviated from the symmetric one, the ratio runs away from
$1$ but still remains a finite value. As expected, this slight modification does not change
the desired $F$-theorem behavior.
In the droplet configurations largely deviated from the symmetric one where
the ratio becomes $0$ or $\infty$,
%we found that the variation of the REE has a positive number.
we found that the variation of the REE has still a negative but an infinite slope for
$(\mu_0 l)^2$, which breaks the perturbative expansion.
%This fact is opposite to our intuition and indicates breaking of the $F$-theorem.
%In order to check the validity for breaking of the $F$-theorem, we investigated the
%validity of the dual LLM geometry.
In this large asymmetric case,
actually the dual LLM geometry becomes highly curved so that the dual gravity description
of the mABJM theory is not allowed and we should also be careful in applying
the AdS/CFT correspondence. Due to this reason,
the appearance of the infinite slope does not indicate the breakdown of the $F$-theorem and
non-stationarity of the REE at the UV fixed point.
%the positive slope of the REE
%does not imply the breaking of the $F$-theorem.
In more general droplet configurations,
it is still difficult to say whether the $F$-theorem is still working or not. Even in the
parameter regions allowing the dual LLM geometry, it is not clear that the slope
of the REE is given by a negative number. It would be interesting to clarify the
REE of the general droplet configurations along the RG flow and helpful to understand
the $F$-theorem and the property of the REE further. We leave it as a future work.

%%%%%%%%%%%%%%%%%%%%%%%%%%%%%%%%%%%%%%%%%%

\section*{Acknowledgements}

This work was supported by the Korea Research
Foundation Grant funded by the Korean Government
with Grant No. 2011-0009972 (O.K.),  NRF-2013R1A1A2A10057490 (C.P.), and
NRF-2012R1A1A2004203, 2012-009117, 2012-046278 (H.S.),
and by the World Class University Grant No. R32-10130 (O.K., C.P.). We thank the
APCTP Focus Program 2014, "Aspects of holography", where parts of the work have been performed.
%%%%%%%%%%%%%%%%%%%%%%%%%%%%%%%%%%%%%%%%%%%%%%

%%%%%%%%%%%%%%%%%%%%%%%%%%%%%%%%%%%%%%%%%%%

\appendix

\section{ Cutting minimal surface and mass deformation }
\label{Appendix}

In \cite{Ryu:2006bv,Ryu:2006ef}, the authors suggested a useful method for mass deformation in a bottom-up approach. The idea is to cut off
the tip of the minimal surface in the conformal case, denoted by $u_0$,
where  the cut-off scale is interpreted as a correlation length $\xi $($< u_0$).
According to this idea, the entanglement entropy for the strip is\begin{align}
S_{\text{strip},~\xi}&= 2 \frac{R^2 L }{4 G_N^{(4)}} \int_{\epsilon}^{\xi} du \frac{\sqrt{ (\frac{dw_2}{du})^2 + 1}}{u^2}\label{bt HEE}\\
&=\frac{R^2 L}{2G_N^{(4)}} \left[  \frac{1}{\epsilon} -\frac{1}{\xi}  + \sum_{n=1}^{\infty}  r_n \frac{\xi^{4n-1}}{l^{4n}}  \right] ~.\nonumber
\end{align}
$r_n$'s are numerical values, some of which are
\begin{align}
r_1 = \frac{2 \Gamma \left(\frac{3}{4}\right)^8}{3 \pi ^2},~r_2=\frac{6 \Gamma \left(\frac{3}{4}\right)^{16}}{7 \pi ^4},~r_3=\frac{20 \Gamma \left(\frac{3}{4}\right)^{24}}{11 \pi ^6},~r_4=\frac{14 \Gamma \left(\frac{3}{4}\right)^{32}}{3 \pi ^8}~~.
\end{align}

When the correlation length is very close to $u_0$, one may take another approximation for (\ref{bt HEE}).  If we express the correlation
length, $\xi \equiv  u_0 (1-\delta^2)$,  in terms of a small parameter
$\delta$, then the above entanglement entropy is approximated as
\begin{align}
S_{\text{strip},~\xi} \sim  \frac{R^2 L}{2 G_N^{(4)}  } \left[   \frac{1}{\epsilon }  -\frac{\Gamma \left(\frac{3}{4}\right)^4}{\pi  l}  -\sqrt{\frac{2}{\pi }}\frac{ \Gamma \left(\frac{3}{4}\right)^2}{l}   \delta+O\left(\delta^{3}\right)  \right] \,,
\end{align}
where $l$ of (\ref{preEE}) has been used.
Up to a multiplicative overall factor, comparing this with
(\ref{stripSA}) gives the following expression for $\delta$
\begin{align}
 \delta =\mu_0^2 l^2   \frac{ \pi ^{5/2} \left( 21 D_3-3 D_1^2-35 D_2\right) } {84 \sqrt{2} \Gamma \left(\frac{3}{4}\right)^6} \,.
\end{align}
This allows us to relate the correlation length $\xi$ of the ABJM theory
to the small mass deformation $\mu_0$ in the mABJM theory. In the symmetric configuration (\ref{sym config}),  $\delta$ is reduced to
\begin{align}
 \delta = \mu _0^2 l^2 \frac{\pi ^{5/2}}{3 \sqrt{2} \Gamma \left(\frac{3}{4}\right)^6} \sim 1.21769 \mu _0^2   l^2 \,.
\end{align}

\section{Curvature Scalar at $y=0$}

To figure out the validity of the gauge/gravity duality in the HEE
calculation, we investigate the behavior of the curvature for general
droplets.
%Though we consider the property of REE near the UV fixed point ($r\to
%\infty$), information of droplets at $y\to 0$ is encoded  the leading
%contribution of the small mass expansion. Hyun:2013sf,Kim:2014yca
As discussed in \cite{Hyun:2013sf}, for some cases the geometry near $y\to 0$ limit
is highly curved even in the large $N$ limit.
The results of the HEE for these cases are not reliable.
In the work \cite{Kim:2014yca}, the authors concentrated on the LLM geometries
corresponding to the case of symmetric droplet
represented by a square shaped  Young-diagram.
Here we generalize this case and investigate the behavior of the curvature at $y=0$ in the large $N$ limit.

The curvature scalar at $y=0$ for general droplet is given by
\begin{align}\label{calR}
l_{{\rm P}}^2{\cal R}(x,y)|_{y=0} = \frac1{6\pi^{\frac23}}\, \frac{Q(\hat x)}{
P(\hat x)},
\end{align}
where $\hat x$  is a rescaled dimensionless coordinate,
$\hat x = 2\pi l_{{\rm P}}^3\mu_0\, x$, and
\begin{align}
P&= \left[g' ( g'- 2 g^2)\right]^{\frac73},
\nn \\
Q &= \left[-40 g^4 {g'}^{3} - 8 g^2 {g'}^{4} + 6 {g'}^{5} + 40 g^3 {g'}^{2} g''
- 12 g{g'}^{3} g'' - 4 g^4 {g''}^{2} - 2 g^2 g'{g''}^{2} + {g'}^{2} {g''}^{2}\right].
\end{align}
The curvature scalar is determined by the function $g(\hat x)$,
\begin{align}
g(\hat x) = \frac12\left(\sum_{i=1}^{2j}\frac{(-1)^{i+1}}{\hat x- \hat x_i}
-\sum^{\infty}_{i=2j+1}\frac{(-1)^{i+1}}{\hat x- \hat x_i} \right),
\end{align}
where  $\hat x_{2j}\leq \hat x \leq \hat x_{2j+1}$ in the $j$-th black strip.
The rescaled coordinate originates from the quantization condition of the
four-form flux~\cite{Cheon:2011gv},
\begin{align}
x_{i+1} - x_i = 2\pi l_{{\rm P}}^3\mu_0(\hat x_{i+1} - \hat x_i )=
2\pi l_{{\rm P}}^3\mu_0\, {\mathbb Z},
\end{align}
where ${\mathbb Z}$ represents an integer and hence
$\hat x_i$'s can be set to integers.
Then the number of M2-branes is represented in terms of $\hat x_i$ as
\begin{align}\label{numberM2}
N = \frac12\left(\sum_{i=1}^{\infty} (-1)^{i+1} \hat x_i^2
-\sum_{i=1}^\infty\sum_{j=1}^\infty (-1)^{i+j}\hat x_i\hat x_j\right).
\end{align}
In the Young-diagram representation, $N$ corresponds to the area of a given diagram.

To obtain reliable results from the gauge/gravity duality, the
dimensionless quantity $l_{{\rm P}}^2 {\cal R}$ should be smaller than
$1$ everywhere in the large $N$ limit.
Now we investigate the behavior of $l_{{\rm P}}^2 {\cal R}(x)$ in
two representative cases.
\\

\noindent
(i) $\hat x_1= -a,\,\, \hat x_2 = 0, \,\, \hat x_3=b$,  $\hat x_4=\hat x_5=\cdots = 0$
case:\\
\noindent
In this case $g(\hat x)$ is given by
\begin{align}
g(\hat x ) = \frac12\left(\frac1{\hat x + a} - \frac1{\hat x} - \frac1{\hat x-b}\right).
\end{align}
Eq.~\eqref{numberM2} tells us the relation $N=ab$ with the range
$1\leq a, b \leq N$.
Let us look at the curvature scalar at the boundary of the black and
white droplet and at the middle of the black droplet.
For convenience, let us set $a= \alpha \sqrt{N}$ with a constant $\alpha$
with the range, $\frac1{\sqrt{N}}\le \alpha \le \sqrt{N}$.
For finite value of $\alpha$, the curvature behaves as
$l_{{\rm P}}^2 {\cal R}\sim N^{-\frac13}$ in the large $N$ limit.
On the other hand,  in the large value of $\alpha$, the leading contribution to the
curvature scalar is given by
\begin{align}\label{recYoung}
&l_{{\rm P}}^2 {\cal R}(0) = \frac13\left(\frac{4\alpha}{\pi\sqrt{N}}\right)^{\frac23},
\nn \\
&l_{{\rm P}}^2 {\cal R}\big(\frac{b}{2}\big) = \left(\frac{2\alpha}{\pi \sqrt{N}}\right)^{\frac23}.
\end{align}
When $\alpha\sim \sqrt{N}$, the curvature scalar is non-vanishing
in the large $N$ limit. That is, we see that the corresponding LLM
geometry becomes
highly curved near $y=0$ and the validity of the gauge/gravity
duality is doubtable.
\\

\noindent
(ii) $\hat x_1= -a-b,\,\, \hat x_2 =-b,\,\, \hat x_3= 0, \,\, \hat x_4=c,\,\, \hat x_5=c+d$
and $\hat x_6=\hat x_7=\cdots = 0$ case:\\
\noindent
Here we set $a= \alpha \sqrt{N}$, $b= \beta \sqrt{N}$, $c=b$, $d=a$ for simplicity.
Then due to the relation \eqref{numberM2}, we have the relation
$\alpha = \sqrt{\beta^2+1}-\beta$ with the range of $\beta$,
$ \frac1{\sqrt{N}} \le \beta \le \frac{\sqrt{N}}{2}$.
When we consider the curvature scalar on the first black droplet, the $g(\hat x)$
is given by
\begin{align}
g(\hat x) = \frac12\left(\frac1{\hat x+a+b} - \frac1{\hat x+b} - \frac1{\hat x} + \frac{1}{\hat x-c} -\frac{1}{\hat x-b-c}\right) \,.
\end{align}
In the small $\beta$ limit near $\hat x=0$, the leading contribution
to the curvature scalar is given by
\begin{align}
l_{{\rm P}}^2{\cal R}(0) = \frac13\left(\frac2{\pi \beta\sqrt{N}}\right)^{\frac23}.
\end{align}
Therefore, we see that when  $\beta\sim \frac1{\sqrt{N}}$ the curvature scalar
is finite in the large $N$ limit.
On the other hand, in the large beta limit, the leading contribution to
the curvature scalar is given by
\begin{align}
l_{{\rm P}}^2{\cal R}(0) =  \frac43\left(\frac{\beta}{\pi \sqrt{N}}\right)^{\frac23}.
\end{align}
We can also obtain the finite curvature scalar in the case of
$\beta\sim \sqrt{N}$ in the large $N$ limit.

From the above investigation on the behavior of the curvature scalar for general droplet near $y=0$, we conclude that, in order to obtain a
geometry weakly curved everywhere in the large $N$ limit,
the length of each edge in the Young-diagram should be proportional
to $\sqrt{N}$.

%%%%%%%%%%%%%%%%%%%%%%%%%%%%%%%%%%%%%%%%%%%

\end{document}